\documentclass[]{IEEEtran}

\usepackage{algorithm}
\usepackage{algorithmic}

\usepackage{listings}
\usepackage[most]{tcolorbox}

\usepackage{hyperref}
\usepackage{amsfonts}

\usepackage{enumitem}

\begin{document}

\newtcblisting[auto counter]{mylisting}[2][]{sharp corners, 
    fonttitle=\bfseries, colframe=gray, listing only,
    listing options={framesep=0.5cm,breaklines=true,mathescape=true,numbers=left,numbersep=-2pt,tabsize=1,literate={\ \ }{{\ }}1}, 
    title=Algorithm \thetcbcounter: #2, #1}

\lstset{
    keywords={func}
}

\title{Blockchain Enabled Data Marketplace - Design and Challenges}

\author{\IEEEauthorblockN{Prabal Banerjee\IEEEauthorrefmark{1}, Sushmita Ruj\IEEEauthorrefmark{2} \\}
\IEEEauthorblockA{
    \IEEEauthorrefmark{1}\IEEEauthorrefmark{2}Indian Statistical Institute, Kolkata, India \\
    \IEEEauthorrefmark{2} CSIRO, Data61, Australia \\
Email: \IEEEauthorrefmark{1}mail.prabal@gmail.com,
\IEEEauthorrefmark{2}sushmita.ruj@data61.csiro.au}}

\flushbottom
\maketitle

\begin{abstract}
Data is of unprecedented importance today. The most valuable companies of today treat data as a commodity, which they trade and earn revenues. To facilitate such trading, data marketplaces have emerged. Present data marketplaces are inadequate as they fail to satisfy all the desirable properties - fairness, efficiency, security, privacy and adherence to regulations. In this article, we propose a blockchain enabled data marketplace solution that fulfills all required properties. We outline the design, show how to design such a system and discuss the challenges in building a complete data marketplace.   
\end{abstract}

\section*{Motivation} 
In this information age, data is the new oil.
From the biggest online giants providing free access to platforms to companies offering reduced health care premium to employees,
 data is the latent currency. Even information about the population of a country are becoming vital for governance.
To an end user, the trade of her data for services is tacit whereas the companies regularly track and trade user data. A simple example would be a recent deal between WhatsApp and Google giving users option to back up WhatsApp chat data to Google Drive without impacting their storage quota.
To users, this would seem like a good deal, but the fine print says that the backed up data is saved unencrypted. So Google would be able to read user data stored on its servers. 

Apart from the widespread data and service trading between companies and its users, there are specific data marketplaces which facilitate data trade by matching demand with information sources. Information providers or sellers showcase their data to woo potential buyers. Interested buyers search and select the information they want and acquire it in exchange for money. The marketplace (mediator) gets incentive for facilitating the trade and, in some cases, for hosting the data. But as data grow in value, so will the cases of cheating and leaks. A key aspect which the conventional data marketplaces tend to ignore is ensuring fair trade. The involved parties - sellers, buyers and mediators, being strategic players, would often want to collude and cheat in order to gain more money. Also, if the mediator hosts the complete trade, and buyer and seller have no communication, then the mediator is in a position of power. It can equivocate and forge for monetary gains. 

Even if such cases are prevented, there are larger regulatory and privacy concerns related to data sharing. Different countries have different privacy laws and hence trade transactions can be pretty complicated. For example, healthcare related data needs to comply with HIPAA. Also, to avoid misuse, the trade history needs to be monitored over time so as to keep tab over ownership and terms of use. 

To tackle the aforementioned problems, we aim to introduce a blockchain layer to the system, which is a distributed tamper-proof append-only ledger. This would enable us to keep historical record of transactions and hence ownership, allow regulatory bodies to (selectively) monitor trade, ensure fair trade with smart contracts and disincentivize collusion using proper incentive mechanism. 

Use of blockchain for trading physical goods have been studied \cite{Goldfeder2017EscrowBitcoin}. While tracking physical entities as states on the blockchain is a tough problem in itself, it is different than trading data. This is because distribution of data to peers may leak information to parties which is not desirable. Also, keeping enormous amounts of data on the blockchain is not possible because of the severe scalability and usability impact. Control of valuable information on the chain so that only the intended recipient gets access is one of the problems we aim to discuss in this article. We would also like to describe the market design, required properties and how existing technologies like blockchain come together to help create an ideal data marketplace. On top of this, we would also try to identify the major challenges that remain in this space.  

\section*{Requirements} 
Digital marketplaces have been studied for a long time, with \cite{Subramanian2017DecentralizedMarketplaces} or without blockchain \cite{Serban2008THEMARKETPLACE, Lacoste1997SEMPER:Marketplace}. Any marketplace primarily consists of two classes of players participating in the trade - the seller $S$ and the buyer $B$. There can be other middlemen who facilitate the trade, such as a marketmaker $M$, but even with their presence, the trade can be seen as a two step trade between $S$ and $M$ followed by $M$ and $B$. In this article, we concentrate on the data exchange part of a marketplace, i.e., $S$ has data $x$ which it wants to sell for price $p$ and $B$ is interested in buying $x$. We discuss digital trades of this kind. We recognize that a marketplace has many more components like discovery and price negotiation, but we discuss them in future works. Let us enumerate the desirable properties of any data marketplace.

The main features of a data marketplace are:
\begin{enumerate}[wide, labelwidth=!, labelindent=0pt]
    \item \textbf{Fairness:} Before executing a trade, the commodity and the price needs to be agreed upon. In our case, the commodity being data, we can define a predicate or condition $\Phi$ that needs to be satisfied by the data. So, $\Phi$ and price $p$ are decided by $S$ and $B$ before trade. We call the trade fair, if either $B$ receives the data satisfying $\Phi$ AND $S$ receives $p$, or neither receives any of it. Any party should be able to pull out of the trade unilaterally, which ensures the timeliness property. \\
    An example of $\Phi$ would be a hash value check so as to ascertain the correctness of the data received. Imagine a file $F$ being traded where both parties agree that for some cryptographic hash function $H$, $H(F)=h$. Then, $\Phi$ is chosen such that $\Phi(F)=1$ iff $H(F)=h$. Under the assumption that $H$ is collision-resistant, it is hard for a malicious party to find $F'$ such that $H(F')=h$, i.e., the malicious party can cheat the agreement over $\Phi$ with negligible probability.
    \item \textbf{Transparency, Privacy and Security:} The information that is being traded must not get leaked to any other party except $B$, that too only if the trade was successful, which would make the system privacy preserving. There should be public log as to ascertain the ownership of data and assurance that at least through this system, transferring data or ownership without keeping log should not be possible. This property does not capture piracy as anyone can leak data outside the marketplace which may be beyond the control of the honest actors of the system. We also want transparency in terms of pricing where, in a trade between $M$ and $B$, $B$ should know what was the original price, terms and conditions of the trade between $M$ and $S$. \\
    An example would be Facebook collecting data from users and selling it to third parties. If such a public log of information transfer was available, scandals like Facebook-Cambridge Analytica 
    could have been identified much faster, if not prevented. Moreover, any party buying the information could be held accountable later as she had complete knowledge of the terms of use of personal data between Facebook and its users. The security property should ensure only legitimate parties gain access to the data itself. 
    \item \textbf{Regulation:} Different types of data have different regulations that may be related to the parties involved in the trade. For example health data have severe regulations like HIPAA. 
     Most regulation enforcement can be tricky because of trade across geographical boundaries and conflicting laws between countries/regions. Any breach of information is severe so the marketplace should provide features that inspire trust among the actors.
    \item \textbf{Efficiency:} For any practical system to have widespread adoption, efficient implementation should be possible. We would call a data marketplace efficient if proposed protocol has comparable speed and requirement compared to state-of-the-art conventional protocols used today. 
\end{enumerate}

\section*{Design}
\subsection*{Conventional Marketplaces}
In a conventional digital marketplace, the seller and buyer interacted either directly with each other or through the marketmaker. If $S$ and $B$ want to trade directly with each other, they would need a Trusted Third Party (TTP) to guarantee complete fairness \cite{Dashti2010FairExchange}, as two party fair exchange is impossible without a TTP. Such a trade would thus ensure either Property 1 or 2 but not both as fairness and privacy are at odds with such a design. This is because, to guarantee fairness, they would need a TTP who would gain access to the data breaching privacy. The other option is to follow iterative fair exchange protocols which would limit damage to fairness but cannot achieve complete fairness. 
Another option is to use the marketmaker $M$ as the TTP. This would have similar problems because $M$ would gain access to data. Often, the data is kept with $M$ and hence there is inherent trust on $M$ by both $S$ and $B$. The problem with this approach is lack of transparency. There has been reports of fraud where the marketmaker has illegally distributed data or tampered metrics to gain unfairly. Popular examples include a 1.6 billion USD lawsuit against popular audio streaming company Spotify for license violation 
and Facebook misreporting key video metrics.

\subsection*{Marketplaces using Blockchain}
Blockchain is a distributed tamper-proof append-only ledger. First widely used in Bitcoin 
it has grown in popularity ever since and has been used to build multiple practical solutions \cite{Kakushadze2018BlockchainPayments}. Under some honest participation percentage assumptions, it can be thought of as a trusted party which can execute arbitrary functionality because of Turing-complete language supported smart-contracts. 

Apart from designing fair protocols using blockchain \cite{Bentov2014HowProtocols, Goldfeder2017EscrowBitcoin,Choudhuri2017FairnessBoards}, there is some existing literature on using blockchain as a TTP or arbitrator facilitating digital data exchange\cite{Herbaut2017AChains, Subramanian2017DecentralizedMarketplaces}. Approaches like in ZKCP \cite{ZKCSP} use Zero-Knowledge which violate our efficiency property (4) for arbitrary $\Phi$. Other approaches \cite{Delgado-Segura2017ATransactions} release parts of data and hence violate the privacy property (2). Recent work like FairSwap\cite{DziembowskiFairSwap:Goods} addresses all the requirements for a two party case. 

There has been work done on developing a practical data marketplace using blockchain. Platforms like IOTA 
and solutions like DDM \cite{DecentralizedDDM}, Fysical 
provide us ways and means to trade information. Other upcoming works like Streamr
aim to provide rich set of features related to data marketplaces. Lot of literature is available related to data trading \cite{MissierMindTrading,CaoAData}, specially IoT and sensor data. While all of them talk about trading data using blockchain, none of them cover all the properties discussed above. 

\begin{figure*}[]
    \centering
    \includegraphics[width=0.8\linewidth]{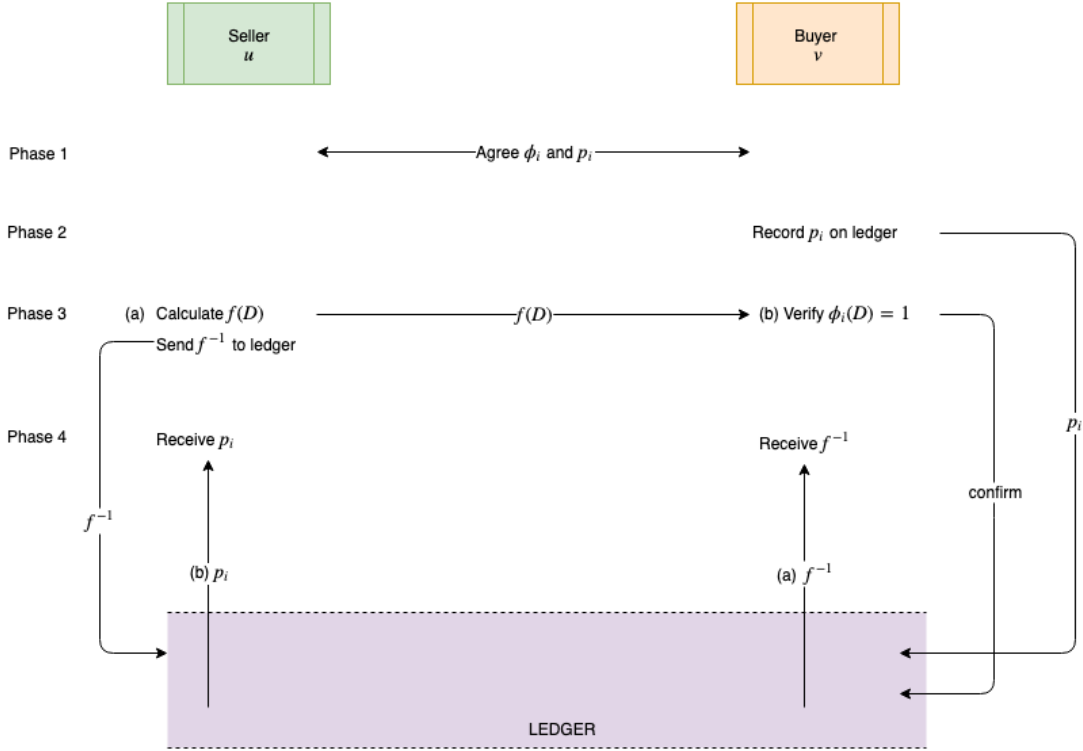}
    \caption{The four phases of a transaction}
    \label{fig:phases}
\end{figure*}

\subsection*{Proposed Design}
Fair Exchange, even between two parties, is impossible without a trusted third party. In our proposed design we want to use blockchain as an enforcer of rules and a trusted third party. It will contain logs of transactions that would help ascertain source and ownership of data being traded, which remains immutable because of the tamper-resistance property of blockchain. At the same time, it would also transfer incentives and act as an automated arbitrator in case of malicious behavior from parties. We use the smart-contract capabilities in the blockchain platform to implement the above functionalities. We assume that the blockchain platform is honest, i.e., the number of malicious actors in the system is below the bound of the platform, and hence can act as a trusted third party facilitating exchange. The proposed design is blockchain platform agnostic as it does not use any feature of the consensus or the communication layer and hence can be implemented on most generic blockchain platforms, as long as it provides Turing-complete programming capabilities. Verifiability comes in-built with a blockchain platform. A database system does not provide validation of a sequence of transactions and hence a blockchain layer is required. We start with a two party data trade case and then go on to describe a three party trade using blockchain as the trusted platform. 

To explain our proposed design, we will use a weighted directed graph to model the interactions where each edge represents the flow of data and vertices represent the different players in the system. 
This graph is not stored as a whole by any of the system participants, but can be constructed entirely or partially as and when required from the information on the ledger. Formally, let $G=(V,E)$ be a directed graph, where $|V| = n$ is the number of players in the system and for each $e \in E$, $e=(u,v)$, $u$ be the seller and $v$ be the buyer for that transaction $e$. We omit the money flow from $v$ to $u$ and we assign weights to edges as the price of data. So, $w(e)=p$ where $p$ is the price of data being sent from $u$ to $v$. 

For each $e_i = (u,v)\in E$, we define $\Phi_i$ to be the condition (or predicate) that $u$ and $v$ agreed upon, according to property 1. $w(e_i)=p_i$ be the price of the data $D$ that is being sold by $u$ and bought by $v$. The players, additionally, have access to a global ledger functionality $\mathbb{L}$, which is the blockchain layer consisting of a smart contract. The internal state of $\mathbb{L}$ is public and it consists of account balances of the parties and contracts. It also provides interfaces to update balances, transfer funds and freeze/unfreeze balances. Freeze transfers funds from a party balance to a contract and unfreeze does the opposite. $\mathbb{L}$ also hosts the smart contract functions which can be triggered using the interface it provides. The function variables are also saved as part of $\mathbb{L}$ internal state.


We outline the process of creation of the graph in Figure \ref{fig:phases}. Initially, let $G=(V,E)$ where $V=E=\phi$. Let $f()$ be an invertible function. For each interested seller $u$ and buyer $v$, the transaction between $u$ and $v$ follows the four phases as given in Figure \ref{fig:phases}. 

We need to note that in Phase 3(b), buyer receives $f(D)$ and not $D$. Hence it should verify another statement $\Phi_i'$ such that $\Phi_i'(f(D))=1 \implies \Phi_i(D)=1$.

If the above protocol terminates successfully, we modify $G$ as $V' = V \cup \{u,v\}$ and $E' = E \cup \{(u,v)\}$. $w((u,v))=p_i$. In case of disputes of over transaction, $\mathbb{L}$ arbitrates and decides whether $\Phi_i(D)=1$. 

As an example, we can think of Alice selling Bob a file $D$ divided into chunks [1, \dots c], for a price of \$5. $f()$ can be a random permutation over 1 to c and $\Phi(D)=1$ iff all chunks contain a particular bitstring $s$. Ofcourse, to the buyer, $f(D)$ is useless but she can verify $\Phi_i(D)=1$ after getting $f(D)$. Later she is revealed $f^{-1}$ which she applies over $f(D)$ to get the desired information. Here, $u$ is Alice, $v$ is Bob, edge is of form $(u,v)$ and $w((u,v)) = 5$. 


To argue that our design can guarantee properties 1-4, we need to give some additional description. 
\begin{itemize}[wide, labelwidth=!, labelindent=0pt]
    \item \textbf{Fairness}: For a single buyer-seller trade, under the assumption that either the seller or the buyer is honest, the design can assure strong fairness guarantees. We can use a protocol like FairSwap~\cite{DziembowskiFairSwap:Goods} to guarantee fairness. The overall idea of FairSwap is that both $B$ and $S$ agree over $\Phi$ which is expressed as a boolean circuit.
    The scheme is designed in such a way that while decrypting if B is unable to decipher information as per agreement, it can generate a Proof of Misbehavior(PoM) and submit to $\mathbb{L}$ which can then punish $S$ for wrongdoing. 
    \item \textbf{Transparency, Security and Privacy}: Every transaction is mediated through $\mathbb{L}$ and hence recorded in the ledger. Any party can ascertain data ownership by looking into the blockchain. For security, we use some encryption function in place of $f$ and some cryptographic hash function in $\Phi$. $f^{-1}$ is the key used to encrypt $D$. For ensuring privacy, we need to ensure that $\mathbb{L}$ does not have access to $D$ but can still verify if $\Phi(D)=1$. This can be done using Zero-Knowledge Proofs or PoM \cite{DziembowskiFairSwap:Goods}. 
    \item \textbf{Regulation}: Checking whether a data trade follows guidelines can be tricky. Here we assume that regulations can be codified and for each trade (edge $e_i$), let $\rho_i$ be the predicate that decides whether the data exchanged $D$ follows the regulations or not. In addition to $\Phi_i$, both buyer and seller inherently agrees to $\rho_i$ when they enter trade. Unlike $\Phi_i$, $\rho_i$ is stored in $\mathbb{L}$ as tamper-resistant log. If some $D$ does not follow $\rho_i$, then it is on the buyer to report the error and cancel the trade, as done for $\Phi_i$. If not, the violations can be found out by law enforcement authorities and penalized on retrospect. 
    \item \textbf{Efficiency}: The major hurdle for efficiency is the complexity of defining and verifying $\Phi_i$ and $\rho_i$. We could use SNARKs, Merkle trees over boolean circuits or state-of-the-art algorithms for analysis over encrypted data. For simple predicates, all of the above techniques are suitable, but there is still work to be done for complex predicates.  
\end{itemize}


\subsection*{Example}
Let us consider an example of such a marketplace. A new online medicine portal wants to give targeted ads and regulate area-wise stock and hence needs medicine consumption and health record data. The data sellers might be small medicine shops in every locality. In practice, however, there will be data aggregators who collect data from local shops and then sends the collated information to the buyer. 

Now let us think of the practical implication of such a trade. The aggregators would have to comply with all health data regulations which can be rather costly for them, whereas their only requirement is to accumulate the information and match the buyer with the sellers. A easier strategy for them would be to just mediate the trade, without keeping the plaintext data. In this case, the sellers transfer encrypted data to the aggregators and send the encryption key directly to the buyer. The aggregator accumulates the data and ensures required predicates are satisfied by all the information collected. The buyer buys the data from the aggregators and sends the money to the blockchain layer which transfers due incentives to all the honest parties. Note that our construction can be extended to capture such interactions and we will describe the details in the next section. 

\section*{Trading Data with Intermediaries}
The last extended example shows that although our design captures most data trades, not necessarily all trade follow our specifications. In particular, we want to have mediators (like the aggregators above) in between sellers and buyers. The previous model assumed any data flow to be a chain of two party trades, but now we want to weaken the assumption and introduce mediators who do not buy or own data, but just facilitate trade with access to only the encrypted data. A typical marketplace hosting website can act as this mediator in a practical scenario. The mediator need not invest in buying the data, but just on the marketplace hosting infrastructure. Like current mediators, say Netflix and Amazon, the mediators may 
perform value-addition by providing exceptional connectivity and advertisement among other services. 
The blockchain smart-contracts ensure that incentive transfer happens automatically from the buyer to the seller and the mediator, as long as the terms of the trade are adhered to. We aim to extend our previous model to capture the intermediaries and show it is still secure.  

With the changed setup, let us state the roles of players in the trade. 
\begin{itemize}[wide, labelwidth=!, labelindent=0pt]
    \item \textbf{Seller} ($S$): $S$ transfers obfuscated information $f(D)$ to a trade mediator and reveals the original information to the buyer in exchange for monetary incentive. She forges information related agreement only with the buyer. The agreement with mediator is to transfer percentage of incentive on successful matchmaking by the mediator.
    \item \textbf{Mediator} ($M$): She collects obfuscated data from $S$ (Potentially collects from multiple sellers and sells the collection to a buyer. But here we consider the simplest case of one seller). She wants to avoid data leak and just wants to earn incentive for successfully facilitating an information transfer. 
    \item \textbf{Buyer} ($B$): $B$ is interested in data $D$ owned by $S$ and wants to acquire the data for some price $p$, given that it satisfies $\Phi$. It gets to know about $D$ and $S$ through $M$ and hence willing to give commission to $M$. 
\end{itemize}

\begin{figure*}[]
    \centering
    \includegraphics[width=0.7\linewidth,height=7cm]{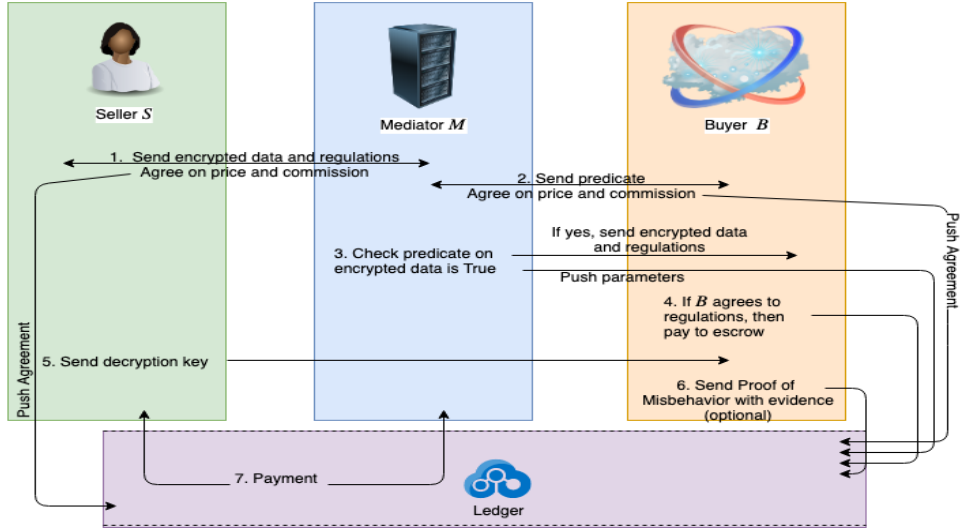}
    \caption{Trading with Intermediaries - Interactions between Parties}
    \label{fig:protocol}
\end{figure*}

The players interact as follows:
\begin{enumerate}[leftmargin=*,font=\bfseries]
    \item $S$ provides $M$ with $f(D), \rho$ and agree on price $p_S$ and commission $c_S \%$. Agreement is sent to $\mathbb{L}$.
    \item $B$ provides $M$ with $\Phi$ and agrees on price $p_B$ and commission $c_B \%$. Agreement is sent to $\mathbb{L}$.
    \item $M$ ensures $\Phi(D)=1$ and sends $f(D), \rho$ to $B$ for price $p=min(p_S,p_B)$. $\rho, p$ and other parameters of the deal is sent to $\mathbb{L}$. 
    \item $B$ checks if it agrees to $\rho$ and proceeds with paying $p+c_B \times p$ to $\mathbb{L}$. 
    \item $S$ sends $f^{-1}$ to $B$.
    \item $B$ sends a complaint with evidence (like PoM) to $\mathbb{L}$ within time $t$.
    \item $\mathbb{L}$ checks complaint and penalizes guilty party as per norms of the system. Otherwise, it sends $(c_S+c_B)\times p$ to $M$ and $p-(c_S)\times p$ to $S$. 
\end{enumerate}

A pictorial representation of the above is given in Figure \ref{fig:protocol}. To capture such mediated trade relationships in our directed graph, we extend $G$ to be $G=(V,E,T)$ where $T$ is a set of three tuples $(s,m,b)$ which represent Buyer, Mediator and Seller respectively. Hence, $s,m,b \in V$. For each tuple, the interaction is as described above. 


\subsection*{Collusion}
 For our original framework, each trade was between two players. As discussed, there exists literature on two party fair trade. But in our extended framework, each trade has three parties. 
  We need to argue that for such interactions, the protocol is resistant to adversarial players and their collusion. We assume here that all players are rational and will act maliciously only for some monetary benefit.
\begin{itemize}[wide, labelwidth=!, labelindent=0pt]
    \item \textbf{Malicious Seller}: A malicious seller $S^{\ast}$ would cheat by sending $D^{\ast}\neq f(D)$ or some $\rho^{\ast} \neq \rho$. Providing $\rho^{\ast}$ makes her vulnerable to retrospective legal actions as a copy of $\rho$ is kept on the chain. For $D^{\ast}$, it would have to lie again in step 5 and hence in step 6 an honest $B$ would raise a complaint and eventually be penalized.
    \item \textbf{Malicious Mediator}: A malicious mediator $M^{\ast}$ may fail to confirm $\Phi(D)=1$ or tamper $f(D)$. For the former, $B$ immediately checks and the deal breaks, hence there is no incentive for $M^{\ast}$ to try that. For the latter, $B$ raises complaint in step 6 and $S$ proves innocence using access to $D$ and agreement in step 1 stored on the blockchain. 
    \item \textbf{Malicious Buyer}: A malicious buyer $B^{\ast}$ may attempt to underpay or raise false complaints. If it directly underpays, it will not receive $f^{-1}$ in step 5, as $S$ checks step 4 before acting in step 5. For false complaints, it should not be able to forge witness. Existing protocols like ZKCP and Proofs of Misbehavior already ensure such properties and hence, with high probability, $B^{\ast}$ will be penalized.
    \item \textbf{Colluding Seller and Mediator}: As both the cases of malicious seller and malicious mediator hinges on an honest buyer, hence this collusion brings no new attack vector. The argument for the two cases collectively works for this situation. 
    \item \textbf{Colluding Seller and Buyer}: $S^{\ast}$ and $B^{\ast}$ may collude to deny $M$ of his commission. The only way to do so is by trading off-chain as the incentive transfer is handled by the smart contract. While using the system, the collusion cannot deny an honest mediator of her commission, under normal honest majority assumptions of the blockchain platform. Our system is not meant to tackle out of system transactions like piracy and hence we do not address the case where the buyer and seller collude to transact outside the system. 
    \item \textbf{Colluding Mediator and Buyer}: $M^{\ast}$ and $B^{\ast}$ might together try to gain access to $D$ without paying $S$. But to do that, $B^{\ast}$ either needs step 5 to be executed before step 4 or lie on step 6. $S$, being honest, will not execute step 5 if step 4 is not complete. False complaint, as argued above, is caught with high probability. Hence, collusion between malicious $M^{\ast}$ and $B^{\ast}$ does not incentivize any of the parties.
\end{itemize}

\subsection*{Example with Intermediaries}
In the previous example, we highlighted the need for an extension to the framework to handle practical challenges that arise during data trade. Let us now look at the actions taken by different parties involved in such a minimal trade. For simplicity, we here assume a single seller, buyer and mediator. 

The local medicine shop (seller) owns data $D$ which contains fields like ($id$, $Age$, $Blood\ Group$, $Class\ of\ Disease$, $Drug\ Name$, $\dots$) where $Class\ of\ Disease\in $ \{$Diabetes$, $Heart\ Ailments$, $Psychological\ Issues$\}. It wants to sell it at a price $p_S$ such that no personally identifiable information is transferred. 
The buyer $B$ wants medical records related to diabetes, heart and psychological drugs sold in the locality. Hence, it describes $\Phi(D)=1$ iff $D$ has a field named $Class\ of\ Disease$ where all entries belong to the set \{$Diabetes$, $Heart\ Ailments$, $Psychological\ Issues$\} and dataset $D$ has at least 10000 rows.

The seller generates a key and encrypts the data using that key, except columns $id$ and $Class\ of\ Disease$. The encrypted data along with signed information on price, commission and other regulation on the data is sent over to the mediator. The buyer sends a signed tuple of requirements and cost with commission that it is willing to pay to the mediator. It receives encrypted data from mediator and checks if requirements match. This check may be done easily in this toy example as parts of data are kept unencrypted, but in reality a primitive like homomorphic encryption or ZKP could be used, as discussed earlier. The buyer also receives the regulations that the data needs to maintain and other pricing information, whose signature it verifies. After matching the details received with that on the ledger, it sends the money to to the ledger after which it receives the key. The key is used to decrypt the data. A complaint is raised if there exists any anomaly in the trade. Again, primitives like, but not limited to, PoM may be used. 

The mediator registers itself before receiving the encrypted data from seller. It also receives the regulation and pricing details, which if it agrees to, it countersigns and sends to the ledger. It receives the requirements from the buyer along with other budget constraints. It checks whether any data matches the budget and information requirements and if so, sends that to the buyer. Again, as earlier, checking property on the encrypted data is easy for this example, but in practice, any of the above discussed primitives may be used. 

\smallskip

The above example pseudocodes can be made arbitrarily complex to cover multiple scenarios like having $\Phi = (\Phi_M$, $\Phi_S)$ where $\Phi_M$ is only checked by Mediator over encrypted data. $\Phi_S$ can contain predicates that the unencrypted data should satisfy and $S$ should terminate trade if $D$ does not satisfy $\Phi_S$. Similarly, the ledger functions can be extended to punish either $M$ or $S$ or both, depending on PoM. Other extensions to the example are multiple sellers using the same mediator, multiple parallel trades processed by same ledger using transaction and player ids. We could have used ZK Proofs instead of PoM. 


\section*{Challenges and Scope of Work}
Let us look at some of the challenges with our proposed design and some areas where there are huge possibilities of improvement.
\begin{enumerate}[wide, labelwidth=!, labelindent=0pt]
    \item \textbf{Global Fairness}: Although we ensured two party fairness using blockchain, we still need to provide fairness guarantees at a much larger context. Looking at a particular information flow, from source to sink, at each vertex it becomes part of a larger dataset. An individual data generator (source) is likely to claim not absolute payment from the next vertex in the flow, but a percentage of incentive from the final data consumer (sink). The percentage can be static, that is fixed during initial sale, or dynamic as the information flows from source to sink through other vertices. Capturing these complicated contracts and tracking the flow for fair payout is a challenge. Each seller will have its own restrictions on how the buyer can use this data. This data changes hands and hence it should also accumulate the usage restrictions provided by its owners along the way. Keeping track of these regulations and its enforcement during trade is a major challenge. If these regulations are kept on blockchain and defined as logical proposition, they can be accumulated as intersection of these propositions. The problem with this kind of approach is that it may make the regulation too strong and hence render the data unusable. Sufficient work needs to be done in this regard. 
    \item \textbf{Efficiency}: Predicate checking over encrypted data for arbitrary logic is still inefficient. Similarly, Zero-Knowledge Proofs which are non-interactive, succinct and captures any complicated generic logic are hard to develop. For example, zk-SNARKs need a trusted setup which can be difficult in a blockchain scenario. On the other hand, STARKs do not need trusted setup and is post-quantum resistant, but the proof size is much larger than SNARK. Bulletproofs have a large proof verification time. In case of Proofs of Misbehavior, they are still inefficient for large circuits. For a circuit with $m$ gates, it takes $O(m)$ communication and computation. Practical homomorphic encryption schemes are only efficient for basic mathematical operations. Fully homomorphic encryption schemes are still orders of magnitude slower than their plaintext counterparts. There is a huge scope for development of algorithms in this space. 
    \item  \textbf{Incentivization}: In this article, we had only provided a basic incentive structure. A much more robust payout strategy needs to be developed for a marketplace use case, which needs thorough game theoretic evaluation and simulation to find the optimal incentive for each players to make foul play economically taxing. All optimal payout strategies may not be useful for a marketplace scenario as the platform needs sufficient margin and liquidity to function properly and these practical factors should also be kept in mind while designing. Capturing practical constraints in a game theoretic framework is a challenge in itself. Disincentivizing foul play by dishonest actors and designing a protocol resilient to colluding malicious parties is a challenge and we wish to pursue this as part of future work. 
    \item \textbf{Codifying Law}: Laws and regulations are written in natural languages. They are made subjective in order to capture a large number of potential cases. They can sometimes be hard to fully capture in a programming language. The deliberate ambiguity in natural languages is at odds with the logical preciseness of programming languages. Even if small nuances are missed, they can lead to huge losses and data breaches. Also, because of the immutable nature of contracts, it is hard to update buggy or outdated contracts once deployed. It is important to capture laws into programmable logic and verified using formal verification methods. A major challenge to formal verification in this space is to draft the specifications precisely. Security audit of contracts and using well known audited open source codes like ERC-20 should be encouraged as much as possible.
    \item \textbf{Data Duplication:} Assuring that the buyer is not buying the same data twice is a problem. Today, most digital data bought from online marketplaces are chosen and bought on the basis of the description given by the owner of the data or the platform. Our proposed design suffers from the same problem. If we use the predicate $\Phi$ to capture such a property by mentioning the data itself which we do not want to buy again, it risks confidentiality of data. Revealing parts of data itself to the buyer to prevent duplicate buys prevents us from delivering fairness. Ensuring data description matches data is a challenge in itself. Moreover, specifying $\Phi$ in a manner that captures requirements property without violating any of the marketplace deliverable properties is a tough task. We need to explore tools like Convergent Encryption which can provide us a way to avoid duplication without revealing anything about the data. 
    \item \textbf{Beyond Data Trade}: A pragmatic data marketplace should have features much more than just data trade, like search, analytics and computation over data. We discussed only the secure data exchange part, but without search discovery and matching services on the marketplace, it is incomplete. For the pricing, we might use auction on blockchain among the interested parties. As we talk of a privacy preserving and secure portal, we need to include techniques from searchable encryption, encrypted analytics and computation over encrypted data. We should provide verifiable results or guarantees so that no party can cheat another and hence provide a fair platform. Each of these problems are worth pursuing in their own right. Inclusion of these added features to the framework would bring added constraints, but we need to show that the basic marketplace properties are still guaranteed. 
\end{enumerate}

\section*{Conclusion}
In this paper, we have enumerated the properties desired of any practical data marketplace, outlined ways and means to design such a marketplace which facilitates trade. Primarily being an idea paper, the exact instantiation of methods and possible tools to be used are not concretely specified. The system design in its entirety is under construction. Implementation of a prototype is ongoing work through which we aim to further understand implementation issues like scalability and performance. The long term goal is to design and implement a production grade marketplace enabling data trade.

\section*{Author Bio}
\textbf{Banerjee, Prabal}: Prabal Banerjee joined Indian Statistical Institute as a Ph.D. student in 2016. He received his B.Sc. and M.Sc. in Computer Science from St. Xavier's College, Kolkata and Chennai Mathematical Institute, Chennai respectively. He is currently working on integration of data with blockchain.

\textbf{Ruj, Sushmita}: Sushmita Ruj is a Senior Research Scientist at CSIRO Data61, Australia. She is also an Associate Professor at Indian Statistical Institute, Kolkata (On Leave). Her research interests are in Blockchains, Applied Cryptography, Data Privacy.  
She is a senior member of the ACM and IEEE.

\bibliographystyle{IEEEtran}

\bibliography{ref}

\end{document}